# Temperature measurements of fusion plasmas produced by petawatt laser-irradiated $D_2$-$^3$He or $CD_4$-$^3$He clustering gases

W. Bang,[1,a] M. Barbui,[2] A. Bonasera,[2,3] G. Dyer,[1] H. J. Quevedo,[1]
K. Hagel,[2] K. Schmidt,[2] F. Consoli,[4] R. De Angelis,[4] P. Andreoli,[4] E. Gaul,[1] A. C.
Bernstein,[1] M. Donovan,[1] M. Barbarino,[2] S. Kimura,[3] M. Mazzocco,[5] J. Sura,[6] J. B.
Natowitz,[2] and T. Ditmire[1]

[1] Center for High Energy Density Science, C1510, University of Texas at Austin, Austin, TX, 78712, USA
[2] Cyclotron Institute, Texas A&M University, College Station, TX, 77843, USA
[3] LNS-INFN, V.S.Sofia 64, 95123 Catania, Italy
[4] Associazione Euratom - ENEA sulla Fusione, via E. Fermi 45, CP 65-00044 Frascati (Rome), Italy
[5] Physics dept., University of Padova and INFN-Padova, Italy
[6] Heavy Ions Laboratory, University of Warsaw, ul. Pasteura 5a, 02-093 Warszawa, Poland

**Abstract**

Two different methods have been employed to determine the plasma temperature in a laser-cluster fusion experiment on the Texas Petawatt laser. In the first, the temperature was derived from time-of-flight data of deuterium ions ejected from exploding $D_2$ or $CD_4$ clusters. In the second, the temperature was measured from the ratio of the rates of two different nuclear fusion reactions occurring in the plasma at the same time: D(d, $^3$He)n and $^3$He(d, p)$^4$He. The temperatures determined by these two methods agree well, which indicates that: i) The ion energy distribution is not significantly distorted when ions travel in the disassembling plasma; ii) The kinetic energy of deuterium ions, especially the "hottest part" responsible for nuclear fusion, is well described by a near-Maxwellian distribution.

---

[a] Author to whom correspondence should be addressed. Electronic mail: dws223@physics.utexas.edu.

Nuclear fusion from explosions of laser-heated clusters has been an active research topic for over a decade [1-11]. Researchers have used explosions of cryogenically cooled deuterium ($D_2$) cluster targets or near-room-temperature deuterated methane ($CD_4$) cluster targets to drive fusion reactions. In these experiments, a gas of clusters is irradiated by a high intensity femtosecond laser pulse. This produces energetic explosions of the clusters, and a high ion temperature plasma results. DD fusion occurring within this high temperature plasma combined with beam-target fusion between the ejected ions of the cluster and surrounding cold gas leads to a burst of fusion neutrons and protons. This phenomenon has been well explained by the Coulomb explosion model [1-5, 12]. In this model, the atoms are ionized and almost all electrons are entirely removed from the clusters after gaining enough kinetic energy from the laser-cluster interaction [1, 13, 14]. This process occurs in such a short time that the ions can be considered nearly stationary: what remains is a highly charged cluster of ions at liquid density, which promptly explodes by Coulomb repulsion producing the multi-keV ions required to initiate the fusion reactions.

The neutron yields from a deuterium fusion plasma depend quite sensitively on the ion temperature and density [15]. The observation of fusion from the laser-irradiated cluster gas implies an effective ion temperature larger than 10 keV, at a density of $\sim 10^{19}$ cm$^{-3}$. Creating these kinds of conditions in most plasma laboratory environments is very challenging. In this Letter, we report on a direct measurement of the ion temperatures in deuterium and deuterated methane cluster plasmas produced by the irradiation of a clustering gas jet by 150 fs petawatt peak power laser pulses. We find that the effective ion temperature produced can be in excess of 25 keV.

Researchers have often used time-of-flight (TOF) diagnostics [11, 16] to measure the effective ion temperature. However, the temperature as measured by a Faraday cup contains the entire time history of the ions, and the effective temperature responsible for the fusion is not accurately measured since the ion velocity distribution could be modified by additional interactions en route to detection (such as collisions in the cold surrounding gas or plasma space charge forces). In the experiment reported here, we implement a technique that was used in an inertial confinement fusion experiment [17], and use the fusion products themselves to measure the ion temperature of cluster fusion plasmas during the time period over which actual fusion takes place, something that has never been conducted before. For the first time in a petawatt laser experiment, a mixture of $D_2$ or $CD_4$ clusters and $^3$He gas is used to observe D(d, $^3$He)n and $^3$He(d, p)$^4$He fusion reactions simultaneously. We measured the yield of 2.45 MeV neutrons and 14.7 MeV protons produced by these two fusion reactions, respectively. Since the cross sections for the two reactions have different dependences on the plasma temperature, the ratio of neutron and proton yields can uniquely determine the plasma temperature at the critical moments and plasma location when the fusion reactions occur.

In our experiment, the Texas Petawatt laser (TPW) delivered 90–180 J per pulse with 150–270 fs duration [18] to irradiate the clusters. It utilized an f/40 focusing mirror (10 m focal length) to create a large interaction volume with laser intensities sufficient to drive cluster fusion reactions. With this laser power and focusing geometry, an increase in the neutron yield many times that of previous cluster fusion experiments was observed [10]. The experimental setup is shown in Fig. 1. Five EJ-232Q and EJ-200 plastic scintillation detectors measured the neutron yields from DD fusion reactions, all of which were calibrated prior to the experiment [19]. Three of these detectors were located at 1.9 m from the fusion plasma, while the other two were located at 5.0 m from the plasma to increase the dynamic range. Four additional NE213 liquid scintillation detectors measured the angular distribution of the fusion neutron emission at four different angles.

Three plastic scintillation detectors measured 14.7 MeV proton yields from the $^3$He(d, p)$^4$He fusion reactions. These were absolutely calibrated prior to the experiment at the Cyclotron Institute, Texas A&M University, using a 14.7 MeV proton beam delivered by the K150 Cyclotron. The proton detectors were located in vacuum 1.06–1.20 m from the plasma at 45º, 90º, and 135º with respect to the laser propagation direction. A 1.10 mm thick aluminum degrader was inserted in front of each detector for two purposes. First, it blocked all the other charged particles including 3 MeV protons from DD fusion reactions. Second, it slowed the 14.7 MeV protons down to 4.0 MeV so that they could transfer all of their remaining kinetic energy to the 254 µm thick BC-400 plastic scintillator disk. When used with 0.25 mm thick aluminum degraders instead, these detectors measured the 3 MeV proton yields. Comparing this with the 2.45 MeV neutron yields allowed cross-calibration of the proton and neutron detectors. The degraders were designed using ion energy loss calculations made with SRIM, a Monte Carlo simulation code [20].

A Faraday cup located 1.07 m away from the fusion plasma collected energetic deuterium and carbon ions arriving from the plasma and provided the ion TOF measurements. The total number of deuterium ions generated at the source was estimated assuming an isotropic emission [16]. This is a legitimate assumption since the cluster expansion dynamics in this experiment belongs to the Coulomb explosion regime rather than the ambipolar expansion regime according to the criteria given in Ref. [4].

We measured pulse energy and pulse duration of the TPW beam for each shot. Two cameras imaged the side and bottom of the plasma during the shot, while a third camera acquired an image that represented the beam profile at the cluster target to estimate the radius, $r$, of the plasma. The measured size of the laser beam was consistent with the side image of the plasma.

Either a cryo-cooled gas mixture of $D_2$ + $^3$He at 86 K or gas mixture of $CD_4$ + $^3$He at 200-260 K served to generate the cluster target. A residual gas analyzer measured the partial pressure of each gas species in the mixture, so the ratio of the atomic number densities of deuterium and $^3$He was known for each shot. The mixtures were introduced at a pressure of 52.5 bars into a conical supersonic nozzle with a throat diameter of 790 µm, an exit radius of $R$ =2.5 mm, and a half angle of 5 degrees to generate large clusters (>10 nm) seeding energetic cluster explosions.

A series of Rayleigh scattering measurements showed that the cluster formation of $D_2$ was not significantly affected (<7%) when $^4$He gas was added, and similar results are expected with the addition of $^3$He into $D_2$ gas. At 86 K, $^3$He atoms do not form clusters [21], and remain cold under laser irradiation because they do not undergo the Coulomb explosions seen by the $D^+$ ions in the clusters. Therefore, in this paper, the "effective plasma temperature" refers to the temperature of deuterium ions only.

With both energetic deuterium ions from Coulomb explosion and cold background $^3$He ions, the possible fusion reactions inside the plasma are:

$$D + D \rightarrow T\ (1.01\ \text{MeV}) + p\ (3.02\ \text{MeV})\ (50\%), \tag{1a}$$

$$D + D \rightarrow {}^3\text{He}\ (0.82\ \text{MeV}) + n\ (2.45\ \text{MeV})\ (50\%), \tag{1b}$$

$$D + {}^3\text{He} \rightarrow {}^4\text{He}\ (3.6\ \text{MeV}) + p\ (14.69\ \text{MeV})\ (100\%), \tag{1c}$$

all of which we could observe with the detectors employed. Figure 2 shows sample results from the Faraday cup, proton detectors, and neutron detectors. The ion TOF data in Fig. 2(a) shows an initial x-ray peak followed by the energetic deuterium ion signal. These data were fitted with good agreement by an exponential decay to account for the response to x-rays, and a Maxwellian distribution for the energetic ions, yielding a TOF ion temperature, $kT_{TOF}$ =10 keV, and a total number of deuterium ions, $N_{ion}$ =1.1×10$^{16}$. It is important to note the plasma is in a non-equilibrium condition [1], and the experimentally observed near-Maxwellian energy distribution is a consequence of the log-normal cluster size distribution [5, 16]. Figure 2(b) shows the x-ray peak followed by the 14.7 MeV proton signal 28 ns later. The proton yield, $Y_p$, is calculated from the height of the proton signal using the previously mentioned calibration. Although each proton detector consists of a very thin (=254 μm) scintillator, there is a small probability (~0.3%) of detecting 2.45 MeV neutrons as well, and this figure shows those neutrons 45 ns after the x-ray peak. Figure 2(c) shows the initial x-ray peak followed by the 2.45 MeV neutron signal 215 ns later, from whose area the neutron yield, $Y_n$, is determined assuming isotropic emission of DD fusion neutrons. The validity of this assumption relies on the angular distribution measurements from the liquid scintillation detectors. The different locations of the neutron and proton detectors with respect to the plasma are responsible for the different neutron arrival times in Fig. 2(b) and 2(c).

Considering a deuterium fusion plasma with temperature $kT$ and density $n_D$ within a volume $V$, the neutron yield from the fusion reaction in Eq. (1b) can be expressed as [2]:

$$Y_n = \frac{1}{2}\int n_D^2 <\sigma_{D(D,n)\,^3He}v>_{kT} dVdt \sim \frac{1}{2}<\sigma_{D(D,n)\,^3He}v>_{kT} t_d \int n_D^2 dV, \qquad (2)$$

where $<\sigma_{D(D,n)\,^3He}v>_{kT}$ is the fusion reactivity and $t_d$ is the disassembly time of the plasma. Calculating the ion temperature from Eq. (2) requires knowledge of all the other variables, some of which are difficult to measure without introducing large errors. This measurement can be greatly simplified by adding cold background $^3$He gas with a known density ratio of $n_{^3He}/n_D$. Then, the ratio of proton and neutron yield from reaction (1c) and (1b), respectively, is:

$$\frac{Y_p}{Y_n} = \frac{\int n_{^3He} n_D <\sigma_{D^3He}v>_{\frac{3}{5}kT} dVdt}{\frac{1}{2}\int n_D^2 <\sigma_{D(D,n)\,^3He}v>_{kT} dVdt} \sim 2\frac{n_{^3He}}{n_D}\frac{<\sigma_{D^3He}v>_{\frac{3}{5}kT}}{<\sigma_{D(D,n)\,^3He}v>_{kT}}, \qquad (3)$$

where $kT$ is the deuterium ion temperature, $<\sigma_{D^3He}v>_{3kT/5}$ is the D($^3$He, p)$^4$He reactivity at the $3kT/5$ center-of-mass temperature (since $^3$He is at rest), and $<\sigma_{D^3He}v>_{3kT/5}/<\sigma_{D(D,n)\,^3He}v>_{kT}$ is a function of $kT$ only that can be calculated (assuming a Maxwellian distribution) using the known fusion cross sections [15]. By only measuring the fusion product ratio with detectors and the density ratio with a residual gas analyzer, the plasma temperature at the critical moments of fusion can be calculated.

In this experiment, the number of fusion reactions between the deuterium ions in the hot plasma and the cold atoms in the background gas outside of the plasma is comparable to that within the plasma and must be taken into account (see Fig. 1-inset for geometry). The proton yield, neutron yield, and their density-weighted ratio are given by [2, 7]:

$$Y_p = N_{ion} n_{3He} <\sigma_{D^3He}>_{\frac{3}{5}kT} R, \qquad (4a)$$

$$Y_n = N_{ion} n_D [\frac{1}{2} <\sigma_{D(D,n)^3He} v>_{kT} \frac{r}{<v>_{kT}} + <\sigma_{D(D,n)^3He}>_{\frac{1}{2}kT} (R-r)], \qquad (4b)$$

$$\frac{Y_p}{Y_n} \frac{n_D}{n_{3He}} = \frac{<\sigma_{D^3He}>_{\frac{3}{5}kT} R}{[\frac{1}{2} <\sigma_{D(D,n)^3He} v>_{kT} \frac{r}{<v>_{kT}} + <\sigma_{D(D,n)^3He}>_{\frac{1}{2}kT} (R-r)]}, \qquad (4c)$$

where $N_{ion}$ is the total number of energetic deuterium ions in the plasma, $<\sigma_{D(D,n)^3He}>_{kT/2}$ is the average fusion cross section between hot deuterium ions and cold deuterium atoms, $R = 2.5$ mm is the radius of the exit nozzle, $r$ is the radius of the cylindrical plasma, and $<v>_{kT}$ is the mean speed of the hot deuterium ions. A uniform atomic density was assumed throughout the gas jet for both $^3$He and deuterium. In this model, $r$ is used in Eq. (4b) to calculate the beam-beam and the beam-target contributions for the DD fusion. The plasma disassembly time is estimated as $r/<v>_{kT}$, and the beam-target contribution is considered only in a distance $R-r$ outside of the fusion plasma.

Figure 3(a) shows the calculated effective ion temperature from Eq. (4c) as a function of the measured density-weighted ratio of the fusion yields ($\equiv Y_p/Y_n * n_D/n_{3He}$) for $D_2 + ^3$He and $CD_4 + ^3$He mixtures. In general, higher ion temperatures, as high as 28 keV, were observed with $CD_4 + ^3$He mixtures than with $D_2 + ^3$He mixtures. This agrees with previous observations [7, 16], and we believe this enhancement in plasma temperature is a result of two combined effects. First, the calculation of the Hagena parameters [22] implies that the $CD_4$ clusters were bigger than the $D_2$ clusters in this experiment, and, second, the average charge density inside a $CD_4$ cluster is higher than that in a $D_2$ cluster [16]. Both effects lead to potentially more energetic Coulomb explosions for the $CD_4$ clusters. In Fig. 3(a), the data did not rest exactly on the dashed line because $r$ was different on each shot.

Figure 3(b) shows a comparison between the temperature measured from the ratio of fusion yields, $kT_{Fusion}$, and the one calculated using ion TOF data, $kT_{TOF}$. The two independent measurements exhibit similar values, although $kT_{TOF}$ was slightly lower than $kT_{Fusion}$ in many cases as if the ions reaching the Faraday cup had lost some energy. SRIM calculations assuming a 2 mm thick uniform gas jet layer at a density of $2\times10^{18}$ atoms/cm$^3$ show that the passage of the ions through the background gas reduces the temperature by 5–10%. The Faraday cup was capable of measuring deuterium ion temperatures up to ~23 keV, above which the ion signal was not distinguishable from the falling edge of the huge initial x-ray peak.

Figure 4 compares the experimentally measured neutron yield with the expected neutron yield calculated using Eq. (4b) and $kT_{Fusion}$, where we used values of $r$, $R$, $N_{ion}$, and $n_D$ measured for each shot in the calculation. The linear relationship shown in this figure supports $kT_{Fusion}$ as correctly representing the ion temperature at the time when the fusion reactions occurred. Up to $1.9\times10^7$ neutrons were produced with $D_2$ + $^3$He mixtures in a single shot, whereas up to $1.4\times10^7$ neutrons were produced with $CD_4$ + $^3$He mixtures. Despite the higher temperatures achieved with $CD_4$ + $^3$He mixtures, the highest neutron yield was achieved with the $D_2$ + $^3$He mixture because of its much higher ion density.

In summary, we have presented results from experiments in which a petawatt laser irradiating a $CD_4$ cluster + $^3$He mixture has produced a maximum deuterium ion temperature of 28 keV and more than $1\times10^7$ fusion neutrons per shot. By adding $^3$He, we successfully measured the ion temperature at the time of fusion reactions using the ratio of measured fusion yields. Within the experimental errors, this temperature agrees with the temperature measured from the ion TOF data. This agreement indicates that the observed Maxwellian energy distribution of the hot deuterium ions is close to the energy distribution of the ions directly responsible for the fusion reactions in the plasma.

WB would like to acknowledge generous support by the Glenn Focht Memorial Fellowship. This work was supported by NNSA Cooperative Agreement DE-FC52-08NA28512 and the DOE Office of Basic Energy Sciences. The Texas A&M University participation was supported by the US DOE and Robert A. Welch Foundation Grant A0330.

**References**

[1]  V. P. Krainov, and M. B. Smirnov, Physics Reports **370**, 237 (2002).
[2]  T. Ditmire *et al.*, Nature **398**, 489 (1999).
[3]  J. Zweiback *et al.*, Phys. Rev. Lett. **84**, 2634 (2000).
[4]  Y. Kishimoto, T. Masaki, and T. Tajima, Phys. Plasmas **9**, 589 (2002).
[5]  J. Zweiback *et al.*, Phys. Plasmas **9**, 3108 (2002).
[6]  I. Last, and J. Jortner, Phys. Rev. Lett. **87**, 033401 (2001).
[7]  G. Grillon *et al.*, Phys. Rev. Lett. **89**, 065005 (2002).
[8]  F. Peano, R. A. Fonseca, and L. O. Silva, Phys. Rev. Lett. **94**, 033401 (2005).
[9]  J. Davis, G. M. Petrov, and A. L. Velikovich, Phys. Plasmas **13**, 064501 (2006).
[10] W. Bang *et al.*, Phys. Rev. E **87**, 023106 (2013).
[11] H. Y. Lu *et al.*, Phys. Rev. A **80**, 051201 (2009).
[12] H. Li *et al.*, Phys. Rev. A **74**, 023201 (2006).
[13] T. Fennel *et al.*, Reviews of Modern Physics **82**, 1793 (2010).
[14] T. Taguchi, T. M. Antonsen, Jr., and H. M. Milchberg, Phys. Rev. Lett. **92**, 205003 (2004).
[15] H. S. Bosch, and G. M. Hale, Nucl. Fusion **32**, 611 (1992).
[16] K. W. Madison *et al.*, Phys. Plasmas **11**, 270 (2004).
[17] V. W. Slivinsky *et al.*, J. Appl. Phys. **49**, 1106 (1978).
[18] E. W. Gaul *et al.*, Appl. Opt. **49**, 1676 (2010).
[19] W. Bang *et al.*, Rev. Sci. Instrum. **83**, 063504 (2012).
[20] J. F. Ziegler, M. D. Ziegler, and J. P. Biersack, Nucl. Instrum. Methods Phys. Res. B **268**, 1818 (2010).
[21] H. Buchenau *et al.*, The Journal of Chemical Physics **92**, 6875 (1990).
[22] O. F. Hagena, and W. Obert, J. Chem. Phys. **56**, 1793 (1972).

**Figures**

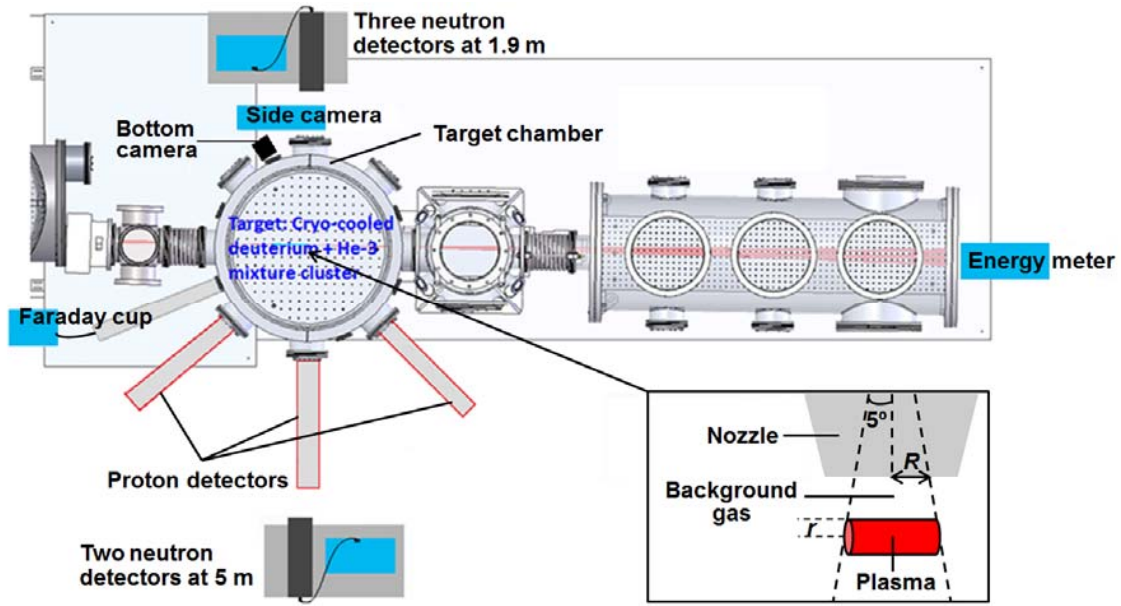

FIG. 1. (Color online) Layout of the experimental area. The laser beam enters from the left, and the nozzle is located near the center of the target chamber. Five neutron detectors and three proton detectors are shown. The inset shows the gas jet nozzle and laser-cluster interaction region.

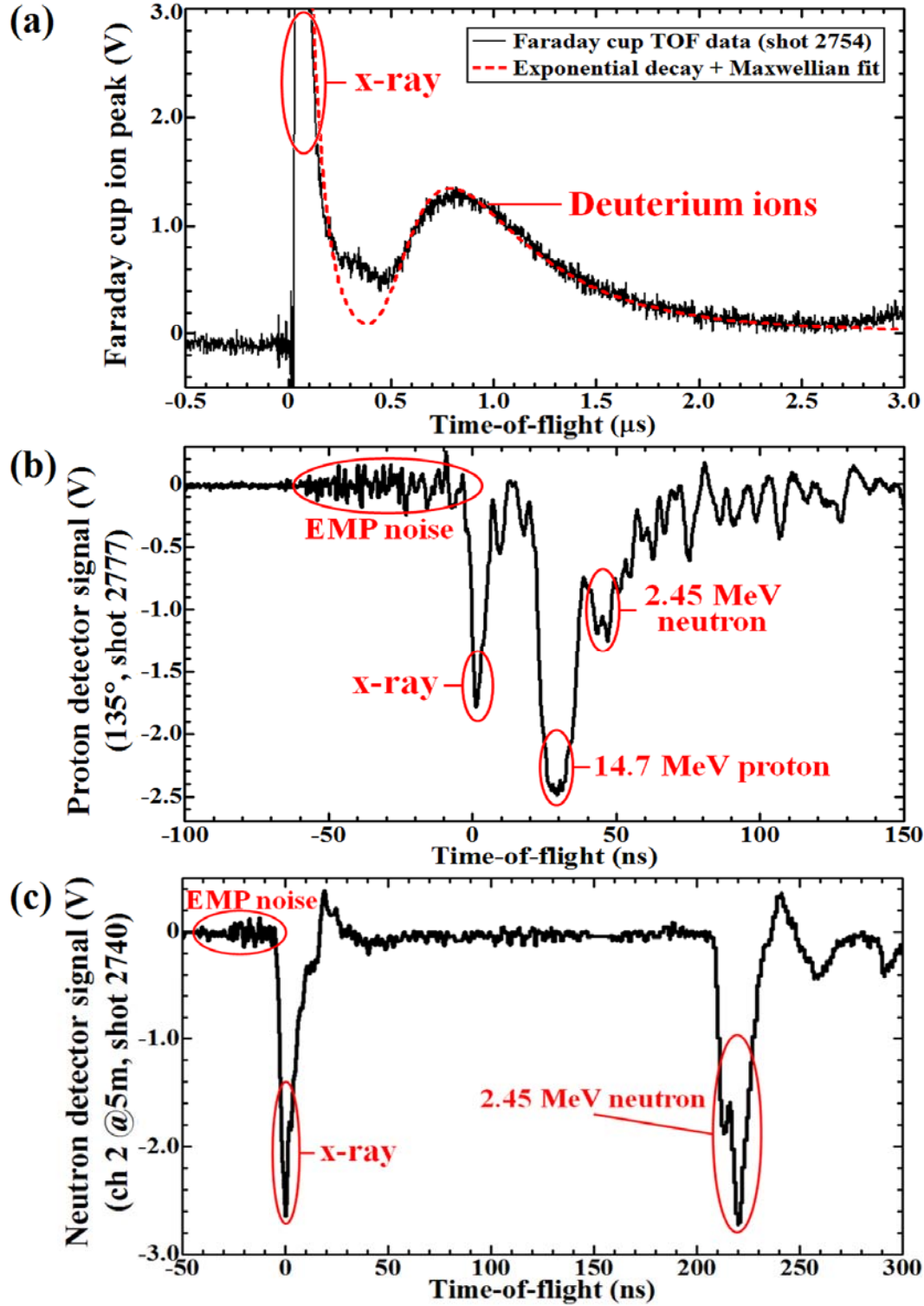

FIG. 2. (Color online) (a) Example of ion time-of-flight data along with a 10 keV Maxwellian fit (dashed red line) and an exponential decay to account for the initial x-ray peak at $t=0$. (b) Proton detector data showing EMP noise, x-ray peak, and the 14.7 MeV proton signal. In this shot, 2.45 MeV neutrons were also detected. (c) Neutron detector data showing EMP noise, x-ray peak, and the 2.45 MeV neutron signal.

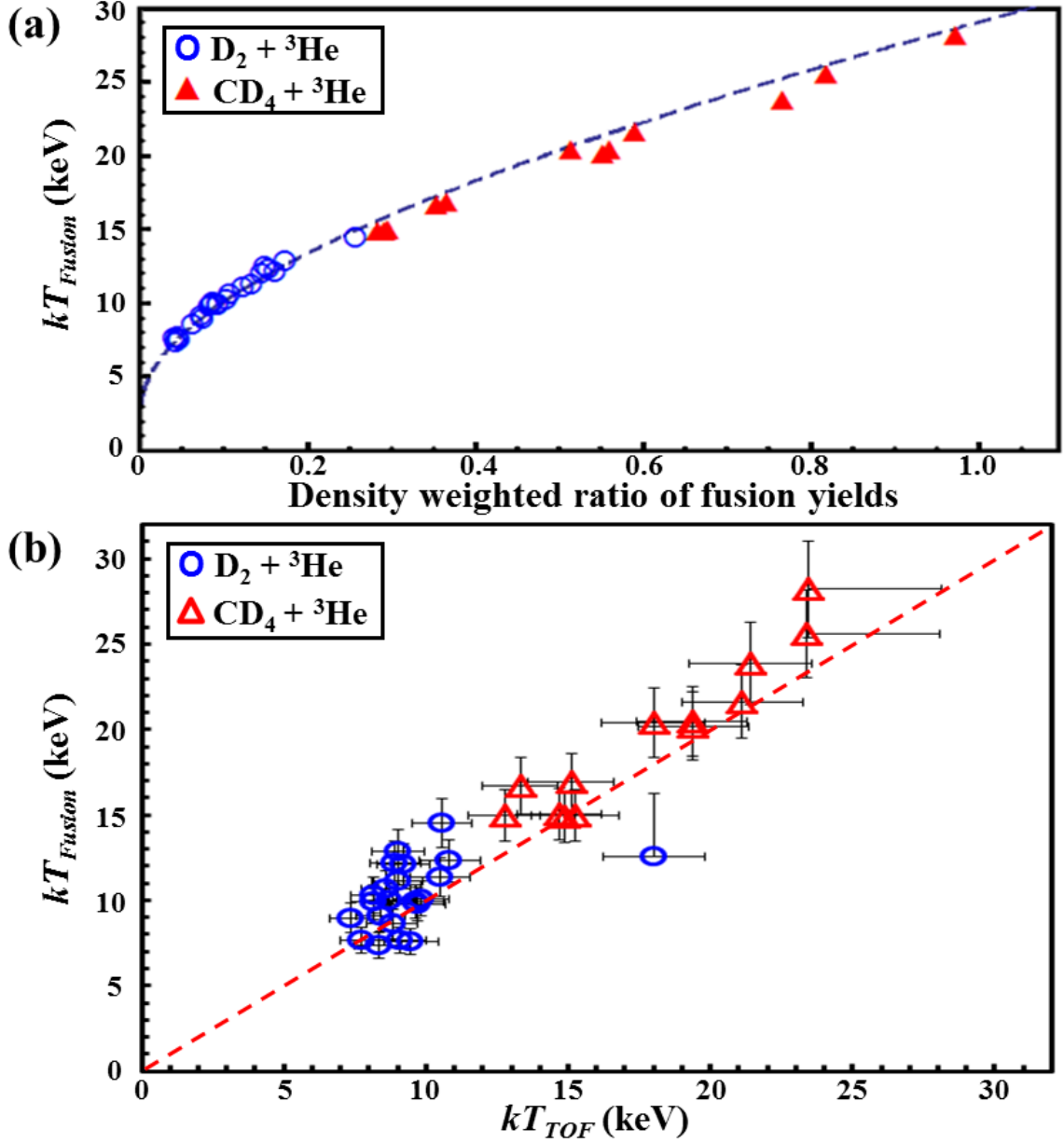

FIG. 3. (Color online) (a) Ion temperature, $kT_{Fusion}$, as a function of the measured density-weighted ratio of fusion yields for $D_2$ + $^3$He (hollow circles) and $CD_4$ + $^3$He (solid triangles) mixtures. The dashed line corresponds to the calculated ion temperature for a plasma size, $r = 250$ μm. (b) Comparison between $kT_{Fusion}$ and the temperature measured by the time-of-flight method, $kT_{TOF}$, for $D_2$ + $^3$He (hollow circles) and $CD_4$ + $^3$He (hollow triangles) mixtures. A dashed line indicates where the temperature determined by both techniques matches.

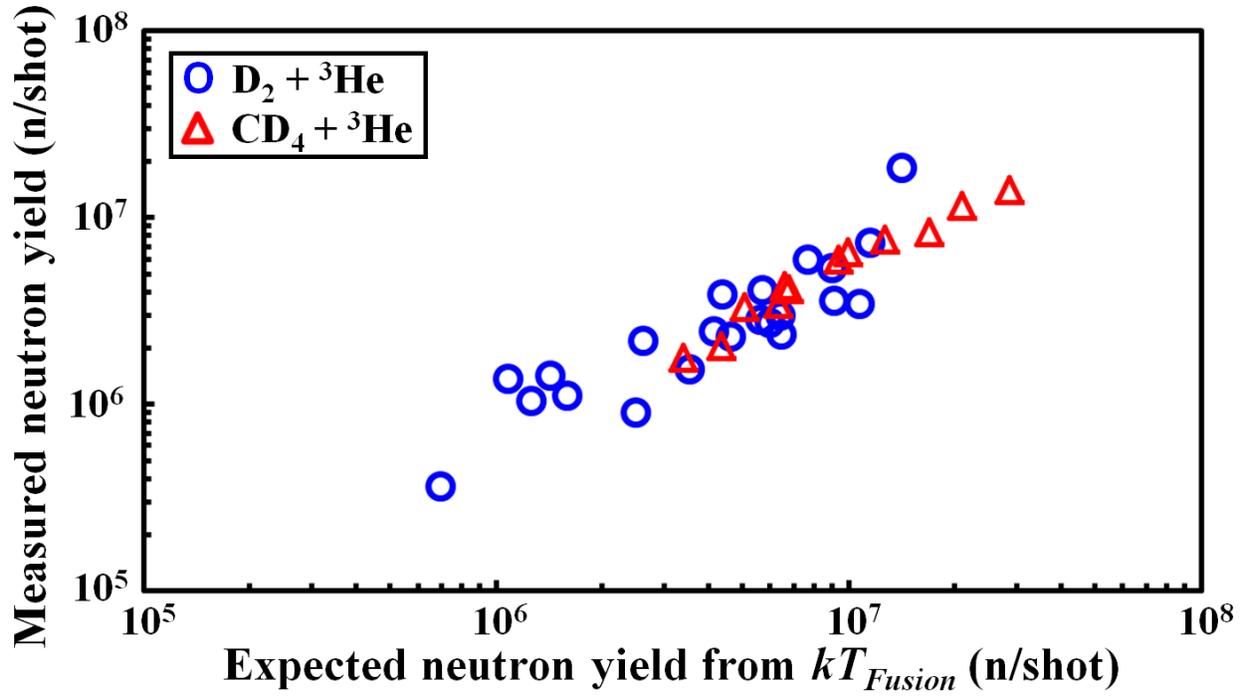

FIG. 4. (Color online) Measured neutron yield versus expected neutron yield from $kT_{Fusion}$.